\documentclass[letter]{aa} 

\usepackage{graphicx}
\usepackage{txfonts}
\begin{document}

   \title{Narrowband oblique whistler-mode waves: Comparing properties observed by
Parker Solar Probe at  <0.3 AU and STEREO at 1 AU}

   \titlerunning{Narrowband oblique whistler-mode waves}

   \author{C. Cattell
          \inst{1}
          \and
          B. Short\inst{1}
          \and
          A. Breneman\inst{1}
          \and
          J. Halekas\inst{2}
          \and
          P. Whittesley\inst{3}
          \and
          D. Larson\inst{3}
          \and
          J. Kasper\inst{4}
          \and
          M. Stevens\inst{5}
          \and
          T. Case\inst{5}
          \and
          M. Moncuquet\inst{9}
          \and
          S. Bale\inst{3,10}
          \and
          J. Bonnell\inst{3}
          \and
          T. Dudok de Wit\inst{6}
          \and
          K. Goetz\inst{1}
          \and
          P. Harvey\inst{3}
          \and
          R. MacDowall\inst{7}
          \and
          D. Malaspina\inst{8,11}
          \and
          M. Maksimovic\inst{9}
          \and
          M. Pulupa\inst{3}
         \and
         K. Goodrich\inst{3} 
         }

   \institute{School of Physics and Astronomy, University of Minnesota, 116 Church St. SE  Minneapolis\\
              \email{cattell@umn.edu}
         \and
             {Department of Physics and Astronomy,University of Iowa, Iowa City, IA 52242, USA}
         \and
            {Space Sciences Laboratory, University of California, Berkeley, CA 94720-7450, USA}
         \and 
             {Climate and Space Sciences and Engineering, University of Michigan, Ann Arbor, MI 48109, USA}
        \and
             {Smithsonian Astrophysical Observatory, Cambridge, MA 02138 USA}
         \and
             {LPC2E, CNRS and University of Orl\'eans, Orl\'eans, France}
         \and
             {Solar System Exploration Division, NASA/Goddard Space Flight Center, Greenbelt, MD, 20771}
         \and
             {Laboratory for Atmospheric and Space Physics, University of Colorado, Boulder, CO 80303, USA}
         \and
            { LESIA, Observatoire de Paris, Universit\'{e} PSL, CNRS, Sorbonne Universit\'{e}, Universit\'{e} de Paris, 5 place Jules Janssen, 92195 Meudon,France}
        \and
            {Department of Physics,
            University of California, Berkeley, Berkeley, CA 94709 USA}
        \and
           {Department of Astrophysical and Planetary Sciences, University of Colorado, Boulder, CO, USA}
            }


 
  \abstract
   {}
   {Large amplitude narrowband obliquely propagating whistler-mode
waves at frequencies of $\sim$0.2 $f_{ce}$(electron cyclotron frequency) are commonly observed
at 1 AU, and are most consistent with the whistler heat flux fan
instability. We want to determine whether similar whistler-mode waves
occur inside 0.3 AU, and how their properties compare to those at 1 AU.}
   {We utilize the waveform capture data from the Parker Solar Probe Fields instrument \textbf{from Encounters 1 through 4} to develop a data base of narrowband whistler
waves. The SWEAP instrument, in conjunction with the quasi-thermal noise
measurement from Fields, provides the electron heat flux, beta, and
other electron parameters.}
   {Parker Solar Probe observations inside $\sim$0.3 AU
show that the waves are often more intermittent than at 1 AU, and interspersed with electrostatic whistler/Bernstein waves at higher
frequencies. This is likely due to the more variable solar wind observed
closer to the Sun. The whistlers usually occur within regions when the
magnetic field is more variable and often with small increases in the solar wind speed. The near-sun
whistler-mode waves are also narrowband and large amplitude, and associated with beta greater than 1.The association with heat flux and beta is
generally consistent with the whistler fan instability. Strong scattering of strahl energy electrons is seen
in association with the waves, providing evidence that the waves
regulate the electron heat flux.}
   {}


   \maketitle


\section{Introduction}

Determining which wave modes control the evolution of solar wind
electrons has long been of interest, from the early studies of their
properties, characterizing three populations -- core, halo and strahl \citep{feldman1975}. Observations indicated that the pitch angle width
of strahl was much broader at 1 AU than would be expected due to the
conservation of the magnetic moment. In addition to collisional
scattering, various wave modes were examined to see if they could
provide the required scattering. Early theoretical work was hampered by 
the lower time resolution measurements of wave spectra~ obtained by
spacecraft in the solar wind. The development of waveform capture
instruments provided high time resolution full waveform data. Studies
utilizing STEREO waveform data near 1 AU revealed the presence of large
amplitude, narrowband whistler-mode waves with frequencies of
$\sim$0.2 $f_{ce}$. The waves propagate at highly
oblique angles to the solar wind magnetic field with significant
parallel electric fields enabling strong interaction with solar wind
electrons without requiring the counter-propagation needed with parallel
propagating waves. These waves are frequently observed, most often in
association with stream interaction regions (SIRs), but also within coronal
mass ejections (CMEs) \citep{breneman2010, cattell2020a} and wave
groups can be observed to last for intervals of days.~

Inside $\sim$0.3 AU, Parker Solar Probe data indicate that
electrostatic waves at higher frequencies ($\sim$0.7 to
several times $f_{ce}$) may be more common \citep{malaspina2020}, 
particularly in regions of quiet radial magnetic field. These
waves include both electron Bernstein and electrostatic whistler-mode
waves. The occurrence frequency decreases with distance from the Sun,
consistent with their absence in the STEREO waveform data at 1 AU. Lower
frequency sunward propagating whistler-mode waves are also observed by
Parker Solar Probe \citep{agapitov2020}, primarily in association
with decreases in the magnetic field or the rapid change in magnetic
field orientation called `switchbacks' or jets \citep{bale2019, kasper2019}.

The properties of the electron distributions have been characterized
inside $\sim$.2 AU by Parker Solar Probe \citep{halekas2020a, halekas2020b}, between $\sim$.3AU and$\sim$.75 AU
by Helios, at 1 AU by Wind and Cluster, and outside 1 AU by Ulysses \citep{maksimovic2005, stverak2009, wilson2019}.
Although the radial dependence of the changes in the properties of core,
halo and strahl are consistent between these studies, the specific
mechanisms that provide the scattering and energization have not been
definitely identified. To understand the role the observed narrow-band
whistler-mode waves play in modifying the electron distributions and
regulating heat flux, it is important to determine how their occurrence
and properties depend on distance from the Sun.

In this report, we describe comparisons of narrowband whistler-mode
waves observed in the waveform data obtained by Parker Solar Probe from
Encounters 1 through 4, and by STEREO. Section 2 presents the data sets
and methodology. Example waveforms and statistical results on the waves
are discussed in Section 3. Conclusions and possible consequences for
solar wind evolution are presented in Section 4.

\section{Data sets and methodology}

We utilize the Level 2 waveform capture data obtained during the first
four solar encounters by the Parker Solar Probe Fields Suite \citep{bale2016}. 
The details of the waveform capture instrument are described
by \citet{malaspina2016}. During the first encounter, three components of
the magnetic field using the search coil instrument were obtained,
enabling determination of the wave vector direction. Subsequent
encounters obtained two components. Although three components of the electric field (potential difference across probes) are transmitted, we utilize primarily the two components in the
plane perpendicular to the spacecraft-Sun line obtained by the longer
antennas. A boom length of 3.5 m is used to covert potential differences to electric fields; a smaller effective boom length would increase electric field amplitudes. The waveform data utilized in this study were obtained for 3.5 s intervals at 150 ksamples/s. As
implemented on STEREO, the highest quality (usually defined by
amplitude of the electric field) captures are stored and transmitted. In addition, intervals of interest in the summary data were selected by the Fields team for transmission of waveform data to the ground. Note that in the first
three encounters dust impacts often triggered the quality flag. For
later encounters, software modifications reduced the number of dust
triggers. The wave amplitudes obtained from the first three encounters
are therefore, on average, smaller than those from the fourth. We also
utilize one electric field and one magnetic field channel in the DC
coupled spectral data, which is obtained at a rate of 1 spectra / 64 Cy,
where 1 Cy = 0.873813 s, over a frequency range of $\sim$10 Hz
to 4.8 kHz \citep{malaspina2016}. The spectra are $\sim$30
s averages. We have also examined one electric field and one magnetic field channel in the DC
coupled bandpass filter (BPF) data which is obtained at a higher cadence of 1 spectrum/ 2 Cy. 

The electron parameters were obtained from the Parker Solar Probe Solar
Wind Electrons Alphas and Protons Investigation (SWEAP) \citep{kasper2016}
 Solar Probe Analyzers (SPAN-A-E and SPAN-B-E) \citep{whittlesey2020}. 
  We utilize electron temperature, temperature anisotropy, heat
flux and density moments and the pitch angle distributions for energies
from 2 to 2000 eV, covering core, halo and strahl \citep{halekas2020a, halekas2020b}.
 The solar wind velocity was obtained from the Level 2
Solar Probe Cup (SPC) moments \citep{case2020}. The solar wind
density, and core and suprathermal electron temperatures were obtained
from the Fields Quasi-thermal Noise (QTN) data \citep{moncuquet2020}.~

\section{Waveform examples and statistics}

\begin{figure*}[h!]
\begin{center}
\includegraphics[width=1\textwidth]{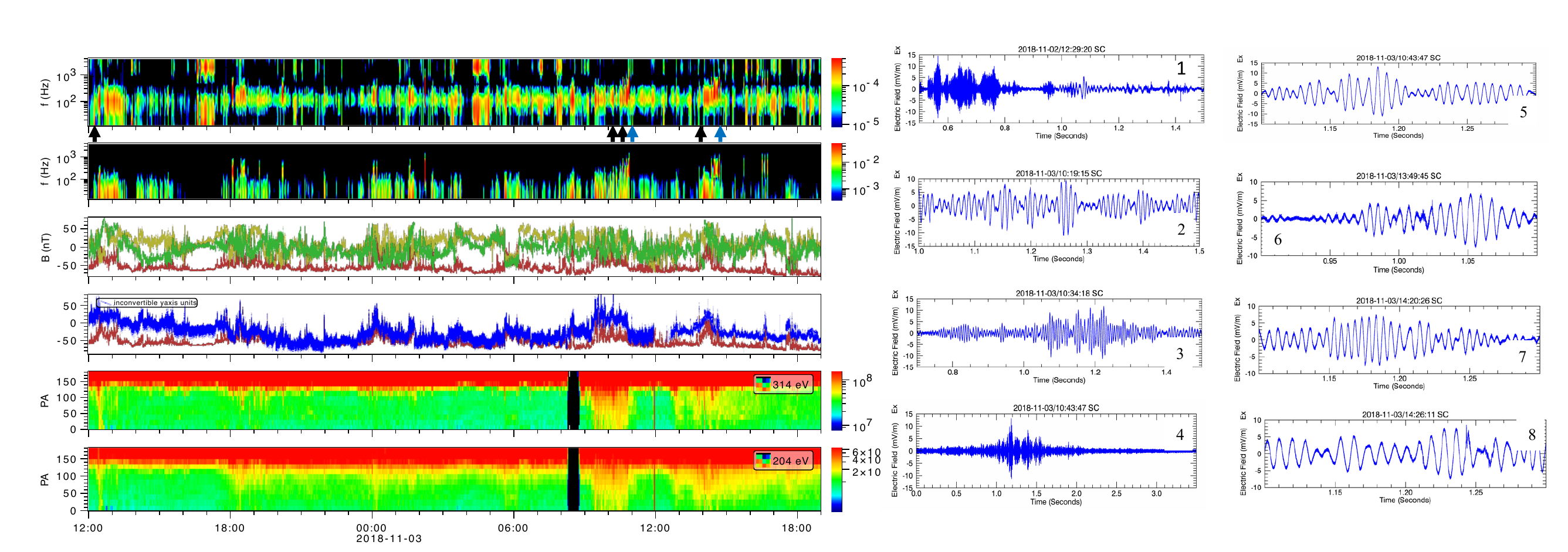}
\caption{{Interval during Encounter 1 with narrow band whistler-mode
waves and higher frequency electrostatic waves. Left panels: DC-coupled BPF
electric field spectrum from 12 to 4000 Hz; DC-coupled BBF magnetic field
from 12 to 4000 Hz; magnetic field in RTN coordinates, R component of magnetic field in red with radial component of ion flow -300 km/s in blue. Pitch angle spectra for electrons with center energy
of 314 and 204 eV. Units for the wave spectra are volts and nT, and for the
electron data are eV/cm\textsuperscript{2}s.~ Right panels: Spacecraft x component of the electric field (in mV/m) snapshots from seven different waveform captures during
this interval at approximate times indicated by arrows with blue arrows indicating more than one snapshot. Note that the
time durations vary. See text for details.
{\label{waveform}}%
}}
\end{center}
\end{figure*}

Figure \ref{waveform} presents an overview of a 31 hour interval from 12 UT on November 2, 2018 to 19 UT on November 3, 2018 that
included 9 waveform captures with narrowband whistlers, as well as
higher frequency electrostatic waves. The top two panels, which plot the
DC-coupled BPF electric field spectrum and the DC-coupled BBF magnetic field
from 12 to 4000 Hz clearly shows the distinction between the higher frequency electrostatic
whistlers/Bernstein waves discussed by \citet{bale2019} and \citet{malaspina2020}
 and the narrowband whistlers that are the focus of this
letter. Examples of the higher frequency electrostatic waves are at
$\sim$1615 to 1715 UT on November 2, and intermittently
between $\sim$03 and 05 UT on November 3, as well as for
shorter intervals on both days. Examples of the narrowband
electromagnetic whistlers can been seen in both spectra at
$\sim$1700 to 1740 UT on November 2, between
$\sim$09 and 11 UT and $\sim$1430 to 15 UT on
November 3, as well as intermittently throughout both days. The fifth and
sixth panels show the pitch angle spectra for electrons with center
energy of 314 and 204 eV, providing evidence for the ability of
these narrow band whistlers to scatter electrons in this energy range.
During the intervals with whistlers, seen in the electric and magnetic
field spectra, there is very significant broadening of the pitch angle
distributions of electrons centered around 314 eV and 204 eV. This
feature is most clearly seen around 1230 and 1800 on 2-11-2020, and
$\sim$930 to 1030 and $\sim$13 to
$\sim$15 on November 3. Note that some changes in the pitch angle distributions are associated with changes in the magnetic field orientation. A detailed discussion of the
scattering and specifics of the resonant mechanisms are presented in \citet{cattell2020b}.
 The fourth panel plots the radial component of the proton plasma velocity in blue (with 300 km/s subtracted to make changes clearer) and the radial component of the magnetic field in red. The third panel plots magnetic field in RTN coordinates. As described in \citet{malaspina2020}, the high frequency electrostatic waves occur primarily in quiet radial magnetic field. The narrowband whistlers occur primarily within regions with
more variable magnetic field and slightly increased flow, and, at times, within or on the edges of structures called
`magnetic switchbacks' or `jets' \citep{bale2019, kasper2019}.

One component of the electric field waveforms for seven of the waveform
captures containing narrowband whistlers is plotted in the right hand set of
panels; \#1 plots 1 s of a waveform captured at 12:29:20 UT on
November 2, showing an interval of high frequency Bernstein waves
followed by whistlers. The rest of the waveforms were observed on
November 3: \#2 shows 0.5 s of a whistler waveform at 10:19:15 UT; \#3
plots 0.8 s of a whistler at 10:34:18 UT; \#4 plots the entire 3.5 s
capture at 10:43:47 UT to show the packet modulations, and \#5 shows the
zoomed in 0.2 s waveform centered on the maximum amplitude; \#6, \#7 and
\#8 plot .2s intervals at 13:49:45 UT, when higher frequency waves were
superimposed on a whistler, at 14:20:26 UT, and at 14:26:11 UT. These
examples show the narrowband coherent nature of the whistler waveforms,
as well as the usual duration of individual sub-packets, and the
amplitude modulation. In the statistics presented below, an event is
defined as a 3.5 s wave capture that contains at least one whistler wave
packet. As these examples show (particularly \#3 and \#4), an event
frequently contains more than one wave packet. Examination of the
magnetic field hodograms (not shown) indicates that the waves are
right-hand polarized, as expected for whistler-mode waves.

\begin{figure}[h!]
\begin{center}
\includegraphics[width=0.85\columnwidth]{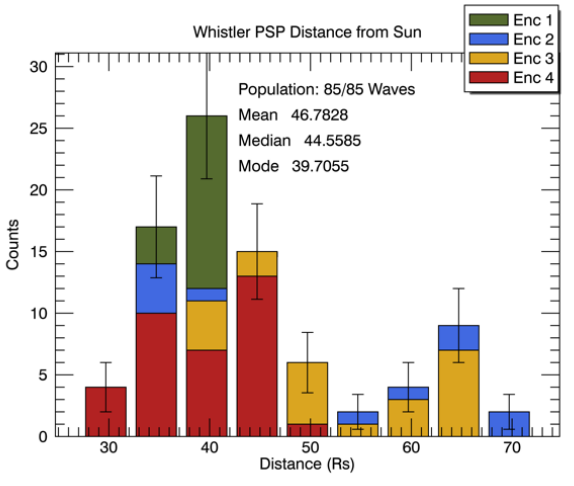}
\caption{Number of narrowband whistler wave captures color coded by encounter number. The number is not normalized by total number of waveform captures obtained.
\label{figure2}
}
\end{center}
\end{figure}

The total number of waveform captures containing narrowband whistlers
versus radial distance is plotted in Figure \ref{figure2}, color coded by encounter
number. Note that instrument modes and solar wind conditions varied between encounters, as did the on-board program for triggering waveform captures. For the 17 waveform captures with whistlers identified in Encounter 1, when three
components of the search coil data were obtained, the wave vector
direction with respect to the background magnetic field and the solar
wind velocity was determined using minimum variance analysis. \textbf{The average wave angle with respect to the magnetic field was 13 degrees, with a maximum of 47 degrees. The average angle is much less than seen in the STEREO data at 1 AU \citep{cattell2020a}; note that there was a very small number of Encounter 1 events compared to the STEREO database for wave angle determination. Comparison of the electric to magnetic field ratio for events seen in the bandpass filter data set suggests that there may be significant number of highly oblique waves. The wave propagation was very oblique to the solar wind velocity. For 13 sunward-propagating events, the average angle to the solar wind velocity was 136 degrees; for the 4 anti-sunward cases the angle was 73 degrees. Phase velocities, ranging from 650 km/s to 1550 km/s, were much larger than the solar wind speed.}

\begin{table*}[h]
\begin{center}
\includegraphics[width=.850\textwidth]{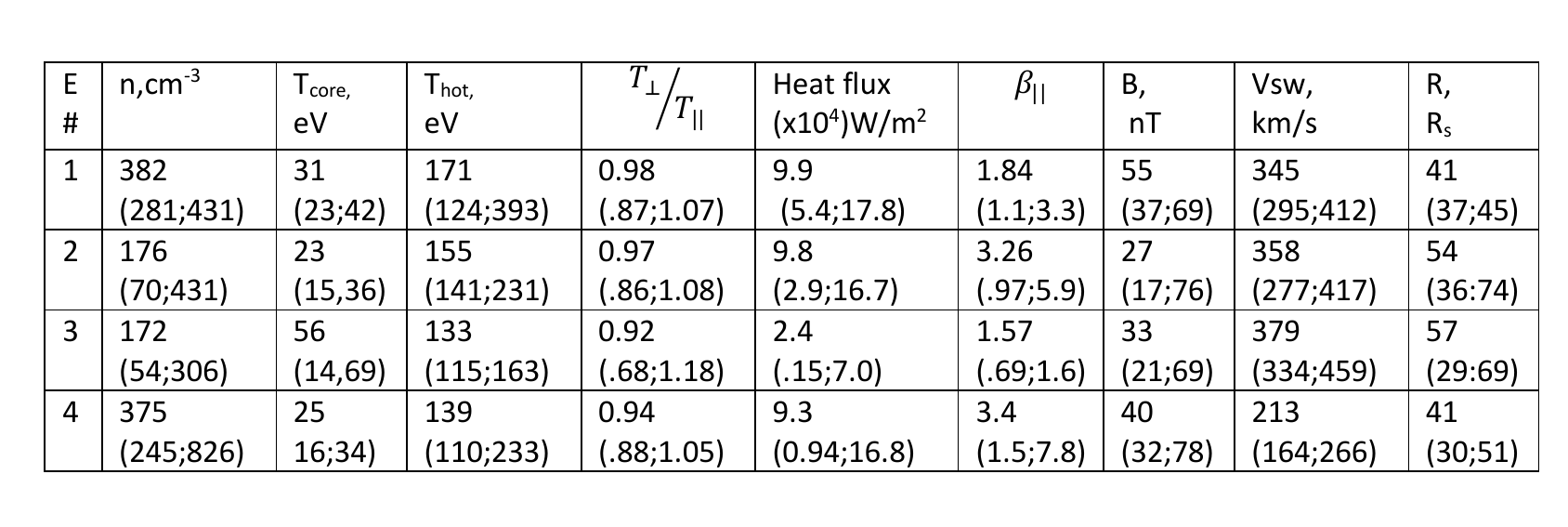}
\caption{\textbf{Average, minimum and maximum values of parameters for Encounters 1 through 4. Average value is the upper values; minimum and maximum are given below in parentheses.} }
\label{table1}
\end{center}
\end{table*}

Statistics of the properties for the waves identified in the first 4 encounters are shown in Figures 3 and 4. The number of events is not normalized by total number of waveform captures obtained. \textbf{Because events in the statistics are color coded by encounter, the range of parameters observed in each encounter is shown in Table 1. The average, minimum and maximum values of the density (cm $^{-3}$), core electron temperature  (eV), suprathermal temperature (eV), temperature anisotropy, heat flux (Watts/m$^2$), $\beta_{e\parallel}$, background magnetic field (nT), solar wind speed (km/s) and distance from the Sun (in solar radii) are shown. There are significant differences in both the average and extreme values between the different encounters, as is also clear in the figures below. Events in Encounters 1 and 4 were on average obtained closer to the Sun in regions of higher density and magnetic field. Events in Encounter 4 were associated with significantly lower solar wind speeds and higher $\beta_{e\parallel}$. Encounter 2 events were associated with the largest core temperature, lowest heat flux and lowest $\beta_{e\parallel}$. Possible explanations for the differences between Encounters 1 and 2 and Encounter 4 were discussed by \citet{halekas2020b}}. 

\begin{figure*}[h!]
\begin{center}
\includegraphics[width=1.00\textwidth]{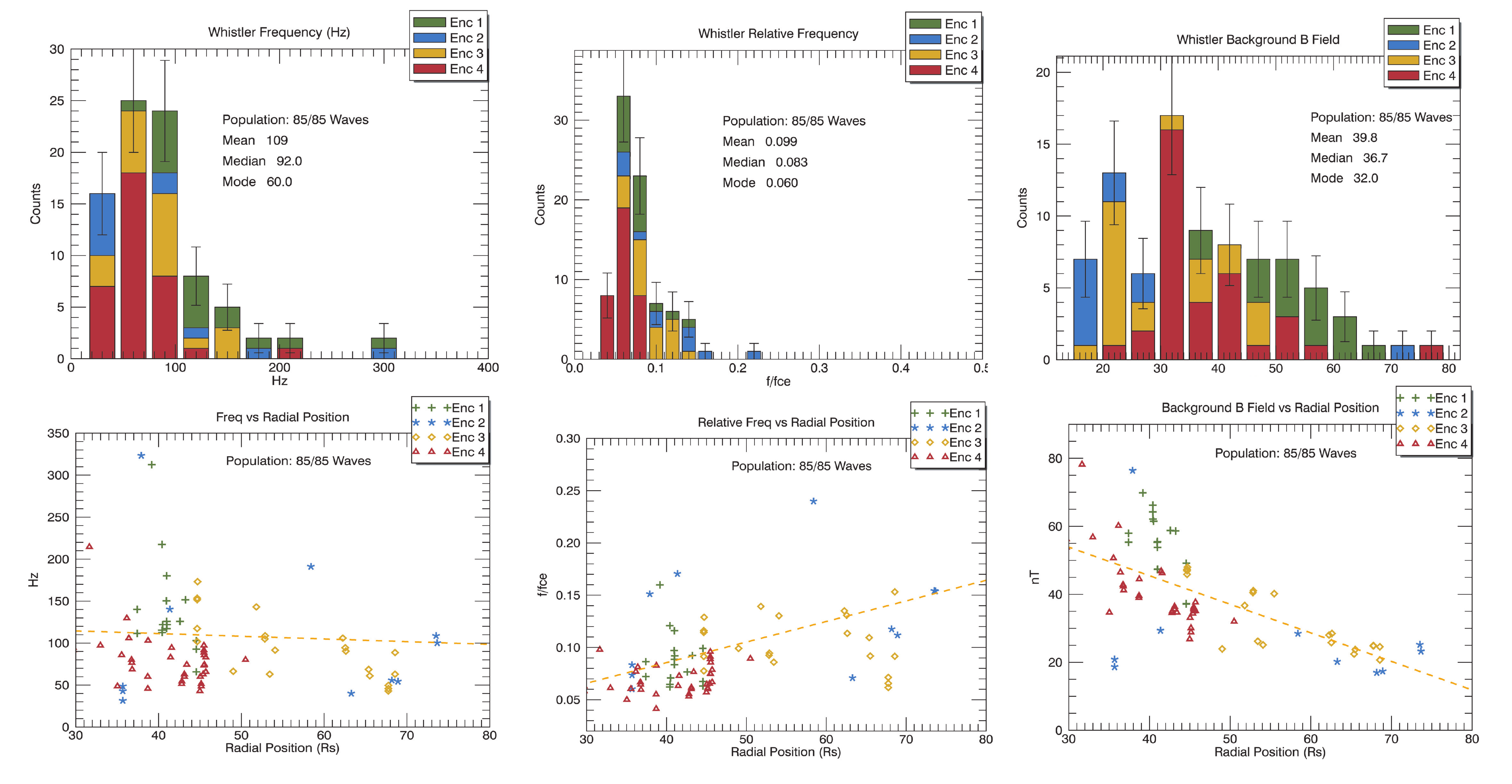}
\caption{Spacecraft frame frequencies for narrowband whistler wave captures color coded by encounter. Top: Number of events versus frequency, frequency normalized by electron cyclotron frequency and magnitude of the background magnetic field.  Bottom: Whistler event frequency, frequency normalized by electron cyclotron frequency and background magnetic field versus radial distance from the Sun.
\label{figure3}
}
\end{center}
\end{figure*}

\begin{figure*}[h!]
\begin{center}
\includegraphics[width=1.00\textwidth]{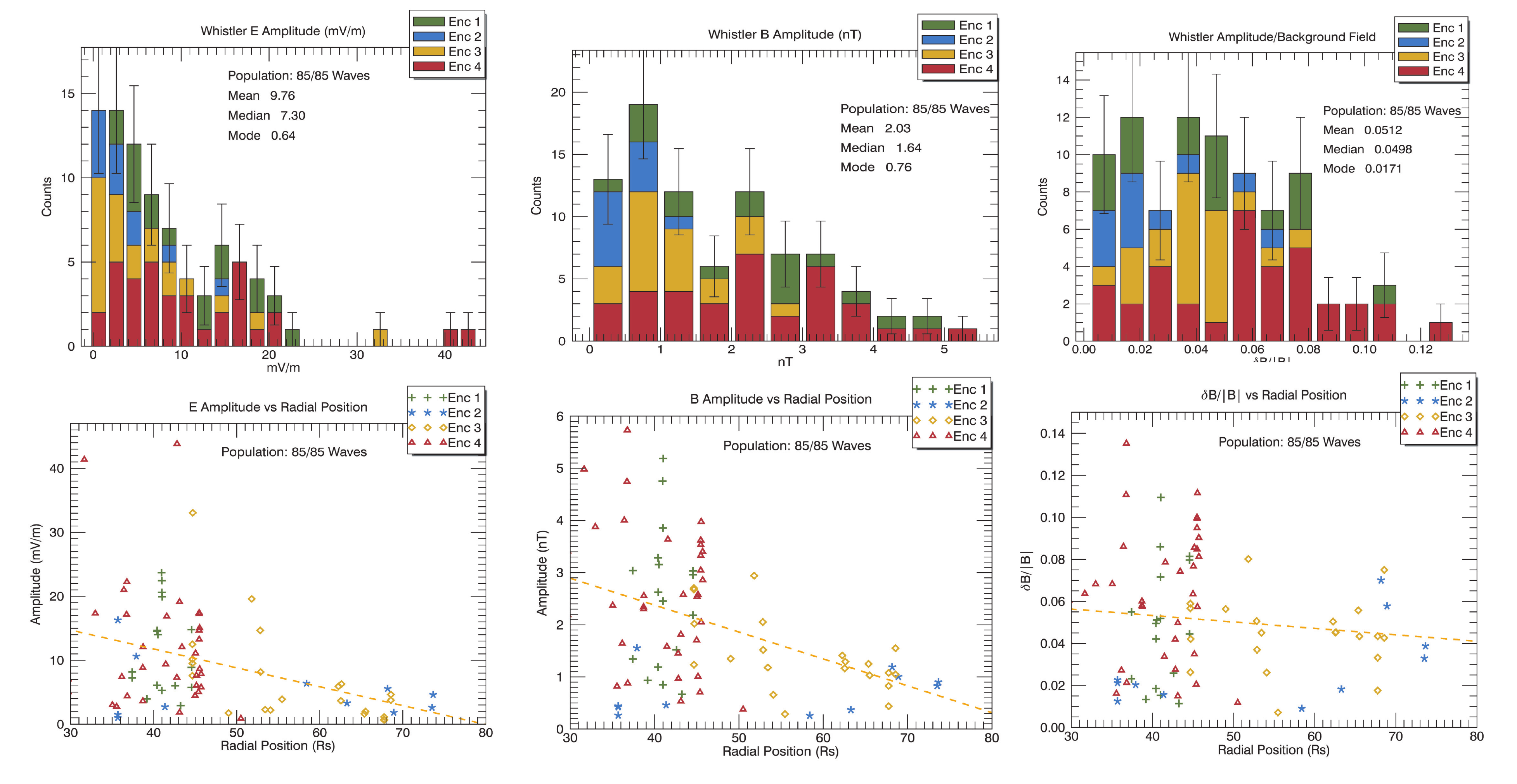}
\caption{Whistler peak amplitudes color coded by encounter. Left
panels: Number of whistler captures versus amplitude of electric field,
magnetic field and magnetic field normalized by background magnetic
field. Right panels:~ Event amplitude of electric field, magnetic field
and magnetic field normalized by background magnetic field versus radial
distance from the Sun.
\label{figure4}
}
\end{center}
\end{figure*}

Figure \ref{figure3} plots the spacecraft
frame frequencies at peak power, color coded by encounter, and the magnitude of the background solar wind magnetic field for each event. The top
panels plot the number of events versus wave frequency, f, and the
number of events versus frequency normalized by electron cyclotron
frequency (f/$f_{ce}$), and the background magnetic field.  The bottom panels plot f,
f/$f_{ce}$ and the magnitude of the solar wind magnetic field versus radial distance from the Sun. There is not
a clear radial dependence of the wave frequency in the spacecraft frame.
In contrast to the case at 1
AU, where \citet{breneman2010} showed that Doppler shifts were
insignificant, there are sometimes significant Doppler shifts in the
waves observed by PSP. \textbf{For the Encounter 1 events, for which the shifts
could be determined, the shifts increased the average
f/$f_{ce}$ to $\sim$0.2, comparable to the value seen at 1 AU \citep{cattell2020a}. The lower frequency whistlers described by \citet{agapitov2020}, utilizing the
lower sample rate fields data set, had significantly larger relative
Doppler shifts. The normalized frequency in the spacecraft frame, f/$f_{ce}$, has a tendency to
increase with distance from the Sun.  Further studies including waveform data from other encounters
will be required to determine the effect of Doppler shifts on the radial dependence.} Whistler events usually occurred in regions with reduced magnetic field magnitudes. 
The wave amplitudes, determined from the peak amplitude seen in any component in each event,
are plotted in Figure \ref{figure4}, color coded by encounter. The top panels show
the number of whistler captures versus amplitude of the wave electric field, amplitude of the wave magnetic field and of the wave magnetic
field normalized by background magnetic field~\(\left(\delta B_w/B_0\right)\)
where~\(\delta B_{w}\) is the magnitude of the wave magnetic field.
The bottom panels show the radial dependence of these amplitudes. There
is a clear decrease in wave amplitudes with radial distance from the
Sun, although the decrease in\(\left(\delta B_w/B_0\right)\) is not as strong. The largest amplitudes were observed close to the Sun during Encounter 4.
Although PSP sees a decrease in the electric field amplitudes with
radial distance, the average amplitude at radial distances around 0.3 AU is only slightly larger than
those observed at 1 AU by STEREO. As noted in Section 2, the amplitudes
for the first three encounters are on average lower than for Encounter 4,
because many waveform captures were triggered by dust until the
algorithm was modified. For this reason, many of the intervals with
whistlers occurred in dust-triggered events rather than ones triggered
by wave amplitude. Data from additional encounters will be required to
determine if the observed amplitude differences between PSP and STEREO
are due to differences in the waveform capture selection criteria or to
physics associated with wave growth and saturation. Note that STEREO did
not have a search coil magnetometer so wave magnetic fields were not
directly measured.

\begin{figure*}[h!]
\begin{center}
\includegraphics[width=1\textwidth]{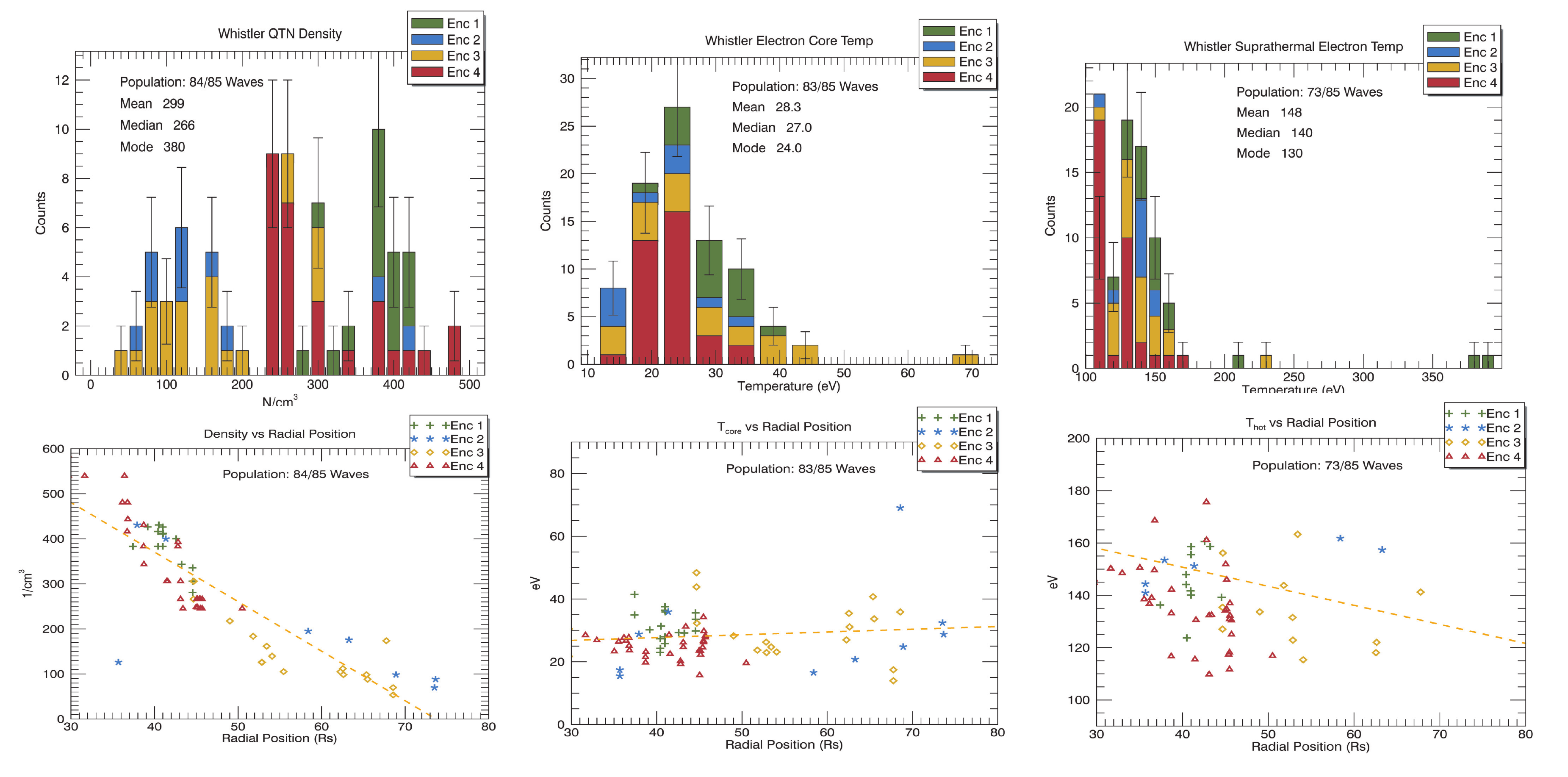}
\caption{Whistler dependence on core density, core and suprathermal
temperature (from the QTN measurement).Top panels plot the number of
events versus core electron density, core and suprathermal temperature.
Bottom panels plot the same quantities versus radial distance from the
Sun.
\label{figure5}
}
\end{center}
\end{figure*}

The association of the whistler events with electron parameters is shown in Figure \ref{figure5}. For most waveform captures, the electron parameters were determined within a few seconds of the capture, with median times of $\sim$ 2 s for QTN parameters and $\sim$ 4 s for SWEAP-determined parameters. For almost all events, the ratio of the electron cyclotron
frequency to the electron plasma frequency (not shown)is
$<$0.01. The left hand panels plot the number of events versus
core electron density, core and suprathermal electron temperature, and
the right panels plot these quantities versus radial distance from the
Sun. These comparisons, which are restricted to events with narrowband
whistler waves, exhibit interesting differences from the results for all
intervals during the encounters 1 and 2 presented by \citet{moncuquet2020}.
 Their results show that the core electron temperature decreases
with radial distance, and the suprathermal temperature was almost
constant. For intervals with the waves, we see a slight increase in the
core temperature, possibly indicating heating of core electrons by the
waves, and a slight decrease in the suprathermal temperature. Note that
our statistics are small and the observed variability at a given radial
distance is as large as the average change with radial distance. 

\begin{figure*}[h!]
\begin{center}
\includegraphics[width=1\textwidth]{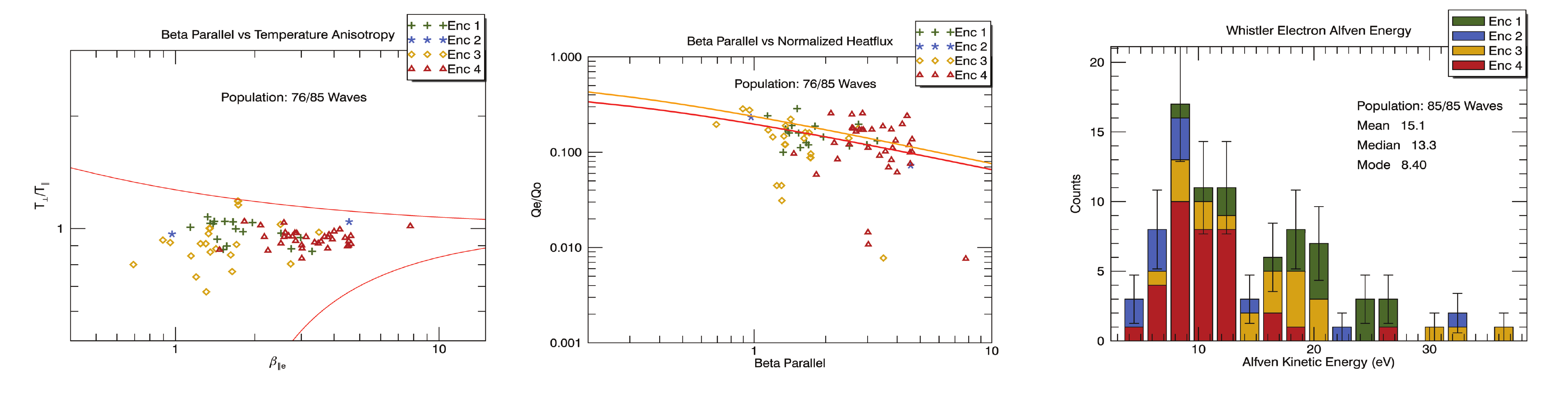}
\caption { \textbf{Comparison to instability mechanisms. From left to right: The temperature anisotropy versus parallel electron beta for wave events. The upper red line is
the whistler temperature anisotropy threshold,\(\frac{T_{e\bot}}{T_{e\parallel}}\) =
1+0.27/\(\beta_{e\parallel}\)\textsuperscript{0.57} and the lower red line is an arbitrary firehose instability (both from Lacombe et al. 2014, based on Gary et al., 1999). Normalized electron heat flux vs \begin{math}\beta_{e\parallel}\end{math}. The yellow and red lines plot the linear instability thresholds for the whistler heat flux fan instability from equation 5 of Vasko et al.(2019) for the parameters of .5 and 1 in their Table 1). The electron Alfven energy to compare to the threshold for the beam mechanism of Sauer and Sydora (2010)}}.
\label{figure6}

\end{center}
\end{figure*}

Possible instability mechanisms are examined in Figures \ref{figure6}. The left panel of Figure \ref{figure6}  shows temperature anisotropy versus parallel electron beta. The
upper red line is the whistler temperature anisotropy
threshold,\(\frac{T_{e\bot}}{T_{e\parallel}}\) =
1+0.27/\(\beta_{e\parallel}\)\textsuperscript{0.57} and the lower red line
is an arbitrary firehose instability (both from \citet{lacombe2014},
based on \citet{gary1999}). The middle panel plots the normalized
electron heat flux versus \begin{math}\beta_{e\parallel}\end{math}\, with the linear instability
threshold for the heat flux fan instability \cite[][for the
parameters of 0.5 and 1 in their Table 1]{vasko2019}. The most striking feature is
that the waves occur when $\beta_{e\parallel}$>1. Many of the largest amplitude waves occurred in Encounter 4, which had significantly higher $\beta_{e\parallel}$.  \citet{halekas2020a}
showed that during encounters 1 and 2 inside 0.24 AU,  $\beta_{e\parallel}$ was usually <1.~ This association of narrowband whistler waves with \begin{math}\beta_{e\parallel}\end{math}
$>$1 was also seen in the STEREO data at 1 AU. The wave
occurrence is constrained by both the whistler temperature anisotropy
threshold and the heat flux fan instability threshold, as was also the
case at 1 AU \citep{cattell2020a}.~\textbf{The temperature anisotropy
destabilizes parallel propagating waves at lower frequencies that we observe.} We conclude, therefore, that the
waves are most likely driven by the fan instability. This is consistent
with the study of the electron heat flux and beta for all intervals
inside 0.25 AU in the first five encounters by \citet{halekas2020b}, and with the conclusion of \citet{agapitov2020} for
a set of events in Encounter 1. Note, however, that there are a significant number
of cases with large $\beta_{e\parallel}$ where the normalized heat flux is \textbf{above the threshold (most from Encounter 4).}

The third panel of Figure \ref{figure6} plots the energy associated with the
electron Alfven speed for the whistler events for comparison to the electron beam driven instability proposed by \citep{sauer2010}.
This mechanism which generates highly oblique waves requires electron
beams that propagate at speeds greater than twice the electron Alfven
speed. The energy associated with the Alfven speed is very low for the
whistler events, $\sim$10 to 20 eV; thus this mechanism would
require beams with energies of $\sim$40 to 80 eV. This is an
order of magnitude lower than would be required for the mechanism to
operate at 1 AU. To date, we have not yet been able to identify beam features at
the appropriate energies in either event list.

To better assess the occurrence probability of these waves we utilized
one electric field channel in the DC coupled
spectral data, at 30 s resolution. We examined by eye the spectral data
for each hour during the first encounter interval shown in the BPF data in Figure \ref{waveform},
which covers 31 hours on November 2 and 3. This yields only a very rough
estimate of occurrence rate. The individual waveform captures (duration 3.5 s) usually contain several individual wave packets. An example of a 3.5 s waveform capture was shown above in Figure 1, packet \#4. The large amplitude whistler packets have durations the order of a few seconds (see \#4, and the shorter duration waveforms in Figure 1), while the lower amplitude waves can last through an entire capture. Waveform \#1 in Figure 1 provides an example when the higher frequency Bernstein waves had the largest amplitude for the initial $\sim$1 s, followed by an interval with both wave types with whistler dominating from $\sim$1 s to 2 s.  We have also have examined the higher cadence BPF data, which can more accurately determine the duration of regions with whistler packets. The BP cadence is comparable to the duration of the observed large amplitude whistler packets. Comparison of the spectral (1 sample /64 Cy) to the BPF data (1 sample/2 Cy) suggests that the occurrence could be on the order of five to 10 percent, based on the electric field spectra.\textbf{There are a significant number of waves that are observed only in the electric field, consistent with very oblique propagation.}  Any definitive determination of an  occurrence rate will depend on the amplitude threshold selected.

\section{Discussion and conclusions}

We have compared statistics of the properties of the narrowband
whistler-mode waves observed in waveform capture data from Parker Solar Probe during the first four
encounters inside $\sim$0.3 AU, to properties observed in waveform capture data from
STEREO at 1 AU. At both radial distances, the waves are narrowband and large amplitude. 
The association with heat flux and beta is generally consistent with the
whistler fan instability. In both data sets the whistlers are observed only for beta >1, and the average temperature anisotropy was $\sim$.9. The PSP electron
data show significant scattering at strahl energies, as documented in
detail by \citet{cattell2020b}. This is consistent with a study of
electron heat flux \citep{halekas2020b} for Encounters 1 through 5, which showed that the heat
flux and beta were constrained by the fan instability threshold,
providing evidence that these waves regulate the electron heat flux.

\textbf{Many instability mechanisms have been proposed for whistler-mode waves in the solar wind that have free energy sources associated with electron properties, including electron temperature anisotropies \citep{gary1996},  heat flux \citep{feldman1975,gary1994} , heat flux fan instability \citep{krafft2010, vasko2019}, the fast magnetosonic/whistler mode \citep{Verscharen_2019}, and electron beam instability \citep{sauer2010}.   Theoretical studies of the dispersion relations have concluded that the most unstable modes for the temperature anisotropy and heat flux instabilities are parallel propagating, and have frequencies f/$f_{ce}$ of $\sim$.01, lower than the frequencies we observe. The transition to oblique modes \citep{gary2011} occurs at values of beta that are much smaller than those observed in the PSP data presented herein.   Only the heat flux fan instability, the magnetosonic/whistler mode, and the beam instability have the highest growth rates at oblique angles.  The PSP whistlers occurred when beta was high, and the events for which Doppler shifts could be determined, \textbf{had frequencies of f/$f_{ce}$ of $\sim$ 0.2. Most cases, however, propagated within 20 degrees of the magnetic field, and none were propagating close to the resonance cone as was seen by  \citet{agapitov2020} in a study of lower frequency whistlers in Encounter 1 and by\citet{cattell2020a} at 1 AU.} The magnetosonic/whistler mode is low beta, and the most unstable modes are at higher frequencies (f/$f_{ce}$ $\sim$.5) than observed at PSP  \citep{Verscharen_2019}. No beam features have been observed in energy range needed for the whistler beam instability. The whistler fan instability is most unstable in the range of f/$f_{ce}$ of $\sim$ 0.1 to 0.2 \citep{vasko2019}, in the range we observe. As shown in Figure 7, the wave occurrence was constrained by the heat fan flux instability threshold.  For these reasons, we conclude that the whistler-mode waves observed by PSP are most likely due to the fan instability, as was also the case for the STEREO whistlers \citep{cattell2020a}. Although the wave occurrence is also constrained by whistler temperature anisotropy, the observed wave properties are not usually consistent with this mode. As discussed below, however, it may be the case that the parallel whistlers and the very oblique whistlers are associated with different instability or saturation mechanisms.} 

\begin{figure*}[h!]
\begin{center}
\includegraphics[width=1\textwidth]{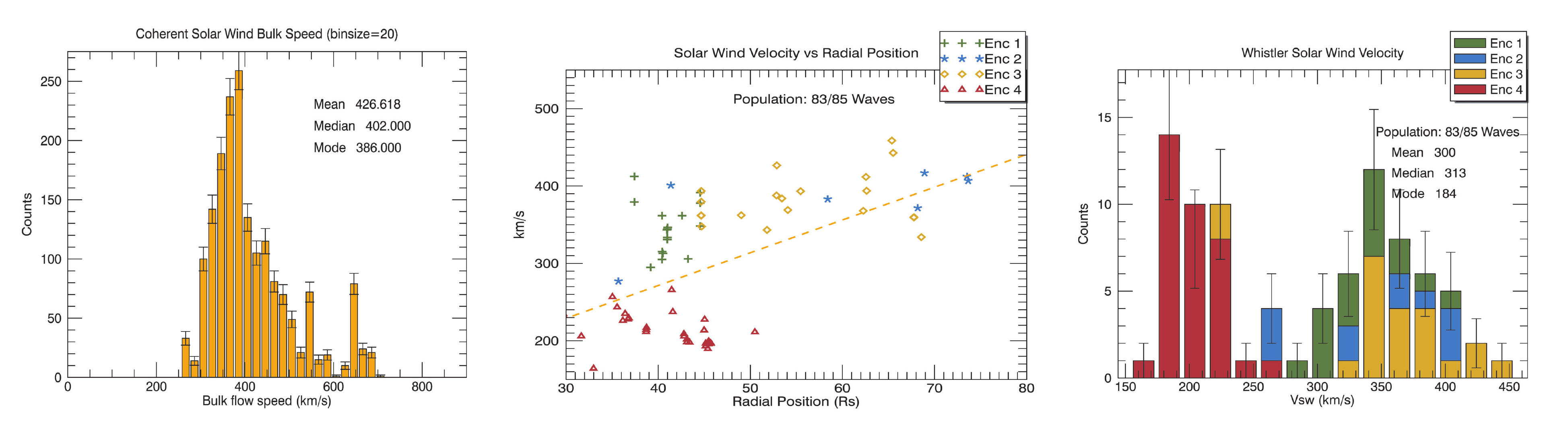}
\caption{The relationship of whistler occurrence to solar wind speed on PSP and STEREO. From left to right: the number of
events versus solar wind speed on STEREO from the \citet{cattell2020a} database, PSP whistler events versus solar wind velocity and radial distance, and  number of PSP whistler  events versus solar wind speed.
\label{figure7}
}
\end{center}
\end{figure*}

At 1 AU, two distinct populations of whistler-mode waves with frequencies of f/$f_{ce}$ of $\sim$0.1 to .2 have been reported from waveform capture data; one population is parallel-propagating with small electric field amplitudes \citep{lacombe2014,graham2017,tong2019} and one is obliquely propagating with resultant large electric fields \citep{breneman2010,cattell2020a}. Although the parallel-propagating waves are usually seen in quiet, slow solar wind \citep{lacombe2014}, \citet{tong2019} show that quasi-parallel whistlers can also occur in the faster solar wind. The oblique waves are often seen in faster solar wind  \citep{breneman2010, cattell2020a}(see Figure \ref{figure7}).   The Parker Solar Probe data shown herein included both parallel and oblique waves, as was also described in \citep{agapitov2020}. There is a tendency for more electrostatic whistlers (i.e. more oblique)to occur within regions of enhanced flow.  A recent study of frequency bank spectral data from Helios \citep{jagarlamudi2020} presented statistics on waves with spacecraft frame frequencies between $\sim f_{lh}$ and 0.5
$f_{ce}$, identified in search coil magnetic field at
distances of 0.3 to 0.9 AU. The observed spectral peaks were identified as whistler-mode based on similarities to \citet{lacombe2014}, but polarization, wave vectors and Doppler shifts could not be determined. The waves were observed almost exclusively in the slow solar wind ($<$400 km/s).  

Figure \ref{figure7} plots histograms of the number of whistler events versus solar wind speed in the \citet{cattell2020a} STEREO database (left) and the number of PSP events color coded by encounter (right) versus solar wind speed. The center panel plots the PSP events versus solar wind speed and radial distance. The highly oblique whistlers observed at 1 AU by STEREO are predominantly seen with solar wind speeds of $\sim$400 km/s, but are also observed with speeds up to $\sim$700 km/s. PSP events are associated with lower solar wind speeds ($\sim$300 km/s). The bi-modal distribution is likely due to to the small number of events, and to radial distance effects, differences in conditions during each encounter, as indicated by the center panel which plots the PSP events versus solar wind speed and radial distance. Encounter 4 events were all obtained inside $\sim$50 solar radii and solar wind speeds were $\sim$200 km/s, whereas events during Encounters 2 and 3 were primarily outside $\sim$50 solar radii with solar wind speeds of $\sim$350 km/s\textbf{(see also Table 1).  The contrast in  solar wind conditions and heat flux during Encounter 4 compared to Encounters 1 and 2 was described by \citet{halekas2020b}. }  The differences in wave association with solar wind speed between PSP events inside .3 AU and the STEREO events at 1 AU may just be due to the evolution of the solar wind. The \citet{jagarlamudi2020} observations which cover the distances between .3 AU and .9 AU, however, were associated with slow flow. Wave vector angles have been determined for only a small fraction of the PSP events; therefore, it is not yet possible to determine if there is a relationship between wave obliquity and solar wind speed at these radial distances.  The parallel propagating waves and the oblique waves may represent two different modes, or different sources of free energy. However, the distinction may also be due differences in instrumentation. Future studies utilizing the Parker Solar Probe data set may resolve the relationship between  the parallel and highly oblique waves.

There are two main differences between the characteristics of the whistlers identified in waveform captures inside
0.2 AU and the waves at 1 AU: (1) the association with larger scale
solar wind properties and structure; and (2) the occurrence of a broader
band less coherent mode at 1 AU that has not yet been identified in PSP
waveform data. In addition, inside 0.2 AU the narrowband electromagnetic
whistlers are interspersed with lower amplitude electrostatic
whistler-mode waves and Bernstein waves at frequencies of
$\sim$0.7 $f_{ce}$ to
$>$$f_{ce}$ \citep{malaspina2020, bale2019}, which have not 
been observed at 1 AU in the STEREO waveform data.

At 1 AU, the narrowband oblique whistlers are most often associated with SIRs, often filling the downstream region of increased
solar wind speed, and often variable magnetic field. The waves are also seen within CMEs
\citep{cattell2020a}.  As shown in Figure \ref{waveform}, inside
$\sim$0.3 AU, the whistlers are associated with intervals of
variable background magnetic field and slight increases in solar wind flow, sometimes
due to magnetic field switchbacks. The association with switchbacks was previously
described by \citet{agapitov2020}. The intervals with packets of narrowband
whistlers can last for several hours, but not for a day or more as seen
at 1 AU. This is most likely due to the much more variable solar wind
conditions observed by PSP close to the Sun. Future studies including
data from additional encounters will examine whether there is an
association of the whistlers with SIRs or CMEs inside
$\sim$.3 AU.

Both the wave magnetic field amplitudes normalized to the background
magnetic field and the electric field amplitudes observed by PSP decrease with radial distance from the Sun, however,  the average electric field amplitudes observed by STEREO at 1 AU
are comparable to those seen by PSP near .3 AU. This may be due to the different selection
criteria for burst data for the two spacecraft or to differences in the
physics. For example, the waves may, on average, be more oblique at 1 AU
or the wave growth and saturation mechanisms may be different due to
differences in the solar wind and electron properties.

Initial results on whistler-mode waves observed by Parker Solar Probe identified in the magnetic field data during the first encounter
were presented by \citet{agapitov2020}, utilizing the
$\sim$300 samples/s waveform data. The observed waves had
large amplitudes ($\sim$2 to 4 nT), often propagated towards the
sun, were significantly Doppler shifted with plasma frame frequencies of
$\sim$0.2 to 0.5 $f_{ce}$, and variable wave
angles. The differences including
average f/$f_{ce}$ warrant additional studies for other
encounters. This will require developing a method to accurately
determine wave vector direction when only two components of the search
coil data are available, utilizing an approach similar to that used on
STEREO based on the three components of the electric field waveform and
the cold plasma dispersion relation \citep{cattell2008)}.

In conclusion, we have shown that narrowband whistler mode waves
observed in the PSP waveform capture data inside $\sim$.3 AU
have many characteristics similar to those seen by STEREO at 1 AU. In
both regions the waves are most consistent with the whistler heat flux fan instability and occur when beta is greater than 1. The waves at 1
AU have slightly higher average f/$f_{ce}$ and are on average
more oblique, but both of these differences may be due to the small
number of PSP events for which we have determined wave angle and Doppler
shifts. When there are wave events, the radial dependence of the core
and suprathermal temperatures is different from that seen for the full electron
data set \citep{halekas2020a}, possibly indicating that the waves heat core electrons. At PSP, the waves are often associated with variable magnetic field and slightly
enhanced solar wind flow, sometimes with `switchbacks,' whereas at 1 AU,
the waves are most often seen in the downstream region of SIRs, also
regions of enhanced flow. Inside $\sim$.3 AU, the regions containing wave packets tend
to last for intervals of hours, whereas at 1 AU, they can last for days.
It is very likely these differences are due to the fact that the solar wind is much more variable on short time scales at PSP compared to at 1
AU.The waves are associated with scattering of strahl energy electrons. A very rough estimate of the wave occurrence at PSP suggests that the waves are often the dominant
wave mode at frequencies below $\sim$3 kHz; combined with the observations of scattering, this suggests that the narrowband whistlers may play a significant role in the evolution of solar wind electrons and regulation of heat flux.

\begin{acknowledgements}
We acknowledge the NASA Parker Solar Probe Mission,
and the FIELDS team led by S. D. Bale, and the SWEAP team led by J. Kasper
for use of data. The FIELDS experiment on the Parker Solar Probe
spacecraft was designed and developed under NASA contract NNN06AA01C.
Work at University of Minnesota and at University of Iowa was supported
under the same contract.\end{acknowledgements}

%
%

\bibliographystyle{aa} 
\bibliography{narrowband} 

\end{document}